\documentclass{article}
\usepackage{graphicx}
\begin{document}

\title{Calibrated Tully-Fisher relations for edge-on galaxies in the BVRI system}
\author {O. Stanchev$^{1}$, P. Nedialkov$^{2}$ and Ts. Georgiev$^{1}$}
\date{March 9, 2004}
\maketitle
$^{1}$\textit{Institute of Astronomy, Bulgarian Academy of Sciences 72, Tsarigradsko Shosse Blvd, 1784 Sofia, Bulgaria}\\
\texttt{stanchev@astro.bas.bg}\\
\texttt{tsgeorg@astro.bas.bg}\\

$^{2}$\textit{Departament of Astronomy, Sofia University 5, James Bourchier Blvd 1164 Sofia, Bulgaria}\\
\texttt{japet@phys.uni-sofia.bg}
\begin{abstract}
Calibrated Tully-Fisher relations ''absolute magnitude - HI line width'' for spiral galaxies oriented edge-on are derived from published data. The slope of the linear fits, especially in R- and I- bands, is close to the value of -10, as predicted by the theory. The standard errors of the relations in B, V, R and I bands are 0.40, 0.36, 0.31 and 0.34 mag, corresponding to relative distance errors of 18\%, 17\%, 14\% and 16\%, respectively. The magnitude of an edge-on galaxy appears to be a good distance indicator.
\end{abstract}

 distance scale;

\section{Introduction}
Tully-Fisher relation (hereafter, TFR) \cite{TF77} between the absolute magnitude $M$ (or the diameter) and HI line width $W$ (or the amplitude of the rotational velocity) of the spiral galaxies is a basic tool to determine galactic distances. The TFR is also used to measure the Hubble constant, to study in details large scale streaming motions etc.

	In the TFR a distance independent observable $W$ ''predicts'' the intrinsic value of distance dependent observable $M$. Further, the difference between the apparent and absolute magnitude $\mu=(m-M)_{0}$ gives directly the distance to the galaxy (in Mpc): $d=10^{(\mu-25)/5}$. The standard error (SE) of the TFR is usually $\sigma_{M}\approx0.5$ mag., corresponding to relative distance error $\sigma_{d}/d=\ln10\sigma_{\mu}/5\approx23\%$. The scatter in the TFR is due to the ''cosmic'' spread of the galactic luminosities with equal $W$.
	
	When TFR is applied to derive the distances, several advantages of the giant spirals are to be pointed out: (a) Because of their high luminosity in optical- and radio-wavelengths they are easily detected at great distances; (b) They avoid the central regions of the clusters, which makes them less affected by the peculiar velocities and more useful for the tracing of large scale flows; (c) In the case of edge-on orientation the inclination corrections to $W$ are negligible and thus the values of $W$ may be adopted without any additionally introduced errors. (d) Because of the high surface brightness gradients of the edge-on galaxies their apparent magnitudes and diameters may be measured with a high accuracy.
	
	The edge-on galaxies are usually excluded from the most of the extragalactic studies because of the uncertain value (possibly too high) of the internal extinction. Nevertheless, the preliminary checks show that the SE of the TFR for the edge-on galaxies is less than in the general case \cite{K89}.  Later the SE of the TFRs for $\sim 3000$ edge-on galaxies from the catalogue \cite{KKK99} was found to be $\sigma \ge 0.5$ mag. \cite{KMK02}. However, these TFRs are not calibrated and the relatively high SE is caused mainly by the use of kinematical (Hubble) distances of the galaxies, containing an additional error due to the galaxy peculiar velocities.

	The question is: How significant is the influence of the internal extinction on the observed magnitudes of the edge-on galaxies, and how good distance indicators they could be? We make use of the most accurate data in four optical bands for 21 nearby galaxies and derive analytically the first calibrated TFRs, referring to the edge-on view of spirals.
	
\section{Data reduction}	
The most representative and precise BVRI data set of 21 nearby galaxies, suitable for calibration of the TFR, is published in \cite{MH00}. In this paper we used 17 galaxies from them. In addition, four galaxies with data taken from RC3 were included in this work. These are the galaxies NGC 224, NGC 598, NGC 2403 and NGC 3031, for which in \cite{SMH00} are no data in V-band.

	Thus our full calibration sample contains 21 galaxies. The magnitudes of these galaxies are recalculated to the edge-on view, in spite of the most common situation where the magnitudes are corrected for internal extinction into face-on view. Reducing the magnitudes to edge-on view, it makes them suitable to calibration of the TFR for target samples of edge-on spiral galaxies.
	
	In order to obtain the edge-on magnitudes, we used the following inclination dependence between the apparent galaxy luminosity $L$ and the inclination corrected galaxy luminosity $L_{0}$ \cite{N94,GHS95}:
\begin{equation}
\log(L_{0}/L)=c_{L}\log(a/b)
\end{equation}
Here $a/b$ is the apparent major-to-minor axis ratio and cL indicates an inclination correction coefficient. In order to perform a calibration of the TFR for edge-on galaxies one need to derive the corresponding edge-on absolute magnitude $M_{e}$. We started with the equation:
\begin{equation}
M_{e}=m_{e}-\mu-A
\end{equation}
In the equation above $\mu$ - is the distance modulus of the calibrator galaxy taken from \cite{SMH00} and $A$ - Galactic extinction from \cite{SFD98}. Expressing equation (1) in magnitudes we derive the following relation for me:
\begin{equation}
m_{e}=m-2.5c_{L}\log\left(\frac{a}{b}q_{0}\right)
\end{equation}
Here $q_{0}$ is the intrinsic axial ratio (or the true compression of the galaxy). This parameter is morphologically dependent and is adopted to be 0.13 for the galaxies of late morphological type ($T>3$) and 0.20 for early types ($T\le3$) \cite{SMH00,TP00}. The empirical constant $c_{L}$ takes into account the influence of the inclination on the magnitude of the galaxy. In the present work we use the values 0.51, 0.40 and 0.34 for $c_{L}$ in B, R and I-band, respectively \cite{N98}, as well 0.46 for V-band, interpolating along the wavelength. 

	Finally, after substituting equation (2) into equation (3), we obtained the expression for the edge-on absolute magnitude of the calibrator galaxies $M_{e}$:
	\begin{equation}
M_{e}=m-2.5c_{L}\log\left(\frac{a}{b}q_{0}\right)-\mu-A
	\end{equation}
Having data for the inclination angle $i$ from \cite{MH00}, we used the Holmberg equation to determine the axis ratio used in the equation above:
	\begin{equation}
\cos^2i=\frac{(b/a)^{2}-q_{0}^{2}}{1-q_{0}^{2}}\Longrightarrow (a/b)=\frac{1}{\sqrt{\cos^2i(1-q_{0}^{2})+q_{0}^{2}}}
	\end{equation}
	With thus derived edge-on magnitudes for each galaxy, the TF calibration relations in BVRI photometric bands were constructed. The differences $\Delta m=(m_{e}-m)$ are given in Table 1. Here $m$ is the galaxy magnitude at arbitrary inclination angle. Apparently, the edge-on magnitudes are weaker by 0.39-0.82 mag. relative to the magnitudes, which are not corrected for inclination.
	
\begin{table}[h]
\begin{center}
\caption{Differences between the edge-on galaxy magnitudes and the magnitudes, corresponding to arbitrary inclination angles}
\begin{tabular}{|c|c|c|c|c|c|}
\hline
Type & Intrinsic   & $\Delta m(B)$ & $\Delta m(V)$ & $\Delta m(R)$ & $\Delta m(I)$\\
Code & axial ratio & $c_{L}=0.51$  & $c_{L}=0.46$  & $c_{L}=0.40$ & $c_{L}=0.37$\\
\hline
$T>3$   & $q_{0}=0.13$ & 0.82 & 0.74 & 0.64 & 0.55\\
$T\le3$ & $q_{0}=0.20$ & 0.58 & 0.52 & 0.45 & 0.39\\
\hline
\end{tabular}
\end{center}
\end{table}

	The basic data on the galaxies are listed in Table 2 as follows: (1) - name of the galaxy, (2) - type code, (3,4,5,6) - the edge-on absolute magnitudes in B, V, R and I bands and (7) - logarithm of HI line width at 20\% of the peak (in km/s). The data for $\log W_{20}$ is taken from \cite{SMH00}.

\begin{table}[ht]
\begin{center}
\caption{The absolute magnitudes and HI line-width parameter for the calibrators}
\begin{tabular}{|c|c|c|c|c|c|c|}
\hline
Name & Type & $M_{B}$ & $M_{V}$ & $M_{R}$ & $M_{I}$ & $\log W_{20}$\\
     & Code &  (mag)  &  (mag)  &  (mag)  & (mag)   &      (km/s)  \\
\hline
1        & 2 &   3   &   4   &   5   &   6   &   7   \\
\hline
NGC 0925 & 3 & 18.82 & 19.21 & 19.59 & 20.03 & 2.4200\\
NGC 1365 & 6 & 20.86 & 21.40 & 21.94 & 22.74 & 2.6820\\
NGC 1425 & 7 & 19.91 & 20.48 & 21.03 & 21.72 & 2.6210\\
NGC 2090 & 3 & 18.77 & 19.27 & 19.83 & 20.62 & 2.5010\\
NGC 2541 & 3 & 17.90 & 18.33 & 18.69 & 19.25 & 2.3700\\
NGC 3198 & 5 & 19.20 & 19.78 & 20.24 & 20.86 & 2.5310\\
NGC 3319 & 6 & 18.55 & 18.98 & 19.31 & 19.80 & 2.4050\\
NGC 3351 & 6 & 18.93 & 19.78 & 20.33 & 21.04 & 2.5860\\
NGC 3368 & 2 & 19.56 & 20.43 & 21.01 & 21.68 & 2.6740\\
NGC 3621 & 5 & 18.66 & 19.24 & 19.83 & 20.41 & 2.4990\\
NGC 3627 & 3 & 20.13 & 20.85 & 21.37 & 22.01 & 2.6260\\
NGC 4414 & 3 & 19.70 & 20.49 & 21.06 & 21.81 & 2.7430\\
NGC 4535 & 2 & 19.68 & 20.37 & 20.74 & 21.53 & 2.5860\\
NGC 4536 & 7 & 19.34 & 20.02 & 20.51 & 21.21 & 2.5620\\
NGC 4548 & 3 & 19.58 & 20.31 & 20.84 & 21.59 & 2.6170\\
NGC 4725 & 5 & 20.20 & 20.89 & 21.39 & 22.06 & 2.6710\\
NGC 7331 & 5 & 20.66 & 21.44 & 22.01 & 22.67 & 2.7460\\
NGC 224  & 4 & 20.12 & 21.03 & 21.53 & 22.25 & 2.7440\\
NGC 598  & 3 & 17.67 & 18.30 & 18.66 & 19.18 & 2.3970\\
NGC 2403 & 2 & 18.15 & 18.47 & 19.16 & 19.67 & 2.4800\\
NGC 3031 & 3 & 19.85 & 20.60 & 21.23 & 21.87 & 2.7190\\
\hline
\end{tabular}
\end{center}
\end{table}
	
	\section{Results}
	TFRs for the calibration sample, based on the data in Table 2, are shown in Figs 1 and 2. Two data sets are placed at each BVRI relation. The filled circles represent the TFRs with the edge-on magnitudes, derived in this work and the open circles show the TFRs with data taken from \cite{SMH00}, obtained through other internal extinction model. The last data set is over plotted for comparison. Solid lines represent the bisector fits, while the dotted lines represent inverse and direct fits, respectively. Dashed line shows the fitting line of the data from \cite{SMH00}. The slopes and zero points relations are determined using bisector fitting technique, which minimize the errors in both $\log W_{20}$ and $M$ axes. It is desirable to apply such a fitting technique for the TFR (see \cite{TF77}), because it is not precisely clear which variable should be treated as the independent variable and which as the dependent one. This complicated problem and the explanation of several fitting methods suitable for the purposes of the extragalactic astronomy is described in detail in \cite{AB96}.
\begin{figure}[h]
\includegraphics[width=12cm,height=7cm]{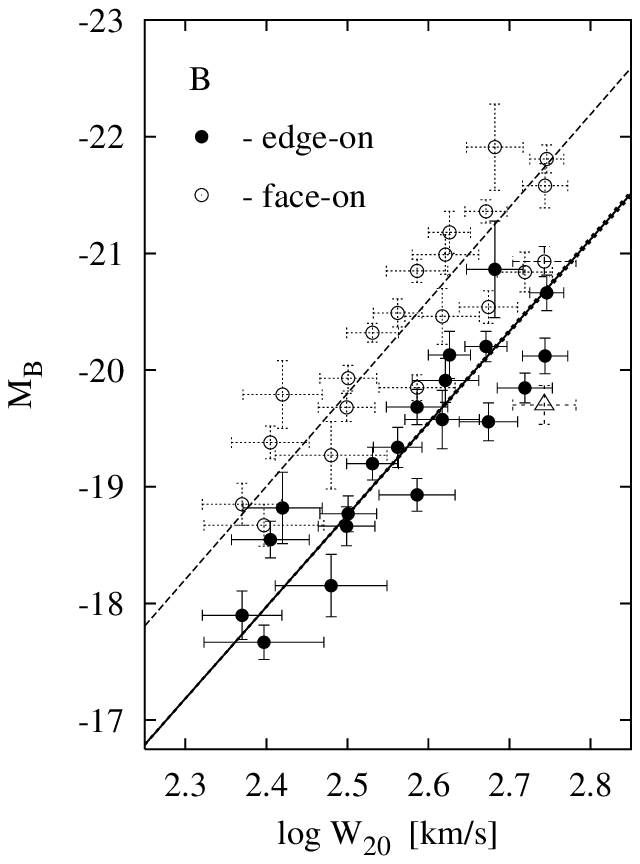}~
\hspace{-6cm}
\includegraphics[width=12cm,height=7cm]{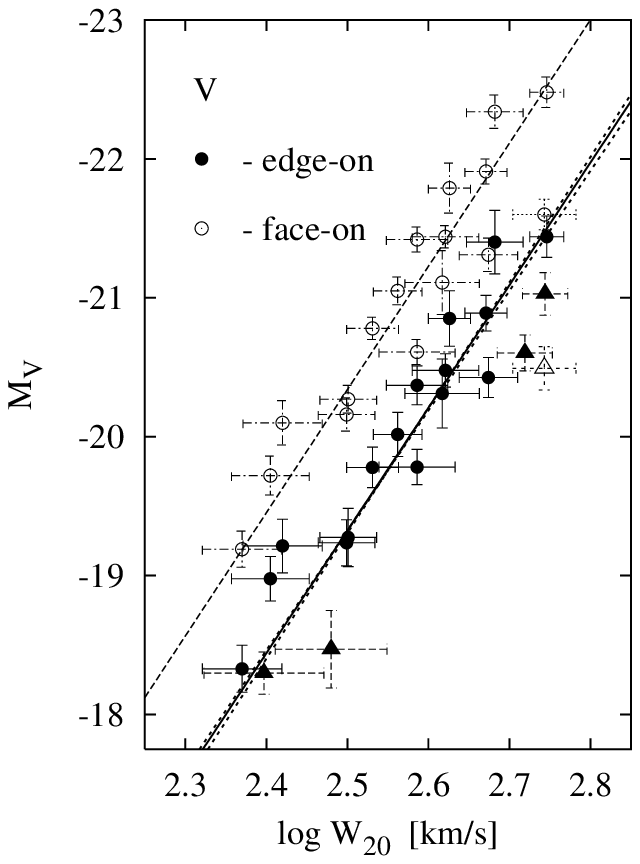}
\centering
\caption{Tully-Fisher relations in B and V-bands. Filled circles represent the relations with the edge-on absolute magnitudes, while the empty circles represent the relation with face-on magnitude data from \cite{SMH00}. The error bars calculated in this work are denoted with solid lines and these taken from \cite{SMH00} - with dashed lines. Solid lines represent the bisector fit. The direct and inverse fits respectively are shown with dashed lines. Dashed line denotes the fit with the data from \cite{SMH00}. One strong deviating galaxy NGC 4414 (denoted with a empty triangle) was not taken into account when determining the fits parameters for the edge-on magnitude relations. The galaxies with RC3 data in V-band are denoted here with fillet triangles.}
\label{Fig. 1.}
\end{figure}
\begin{figure}[h]
\includegraphics[width=12cm,height=7cm]{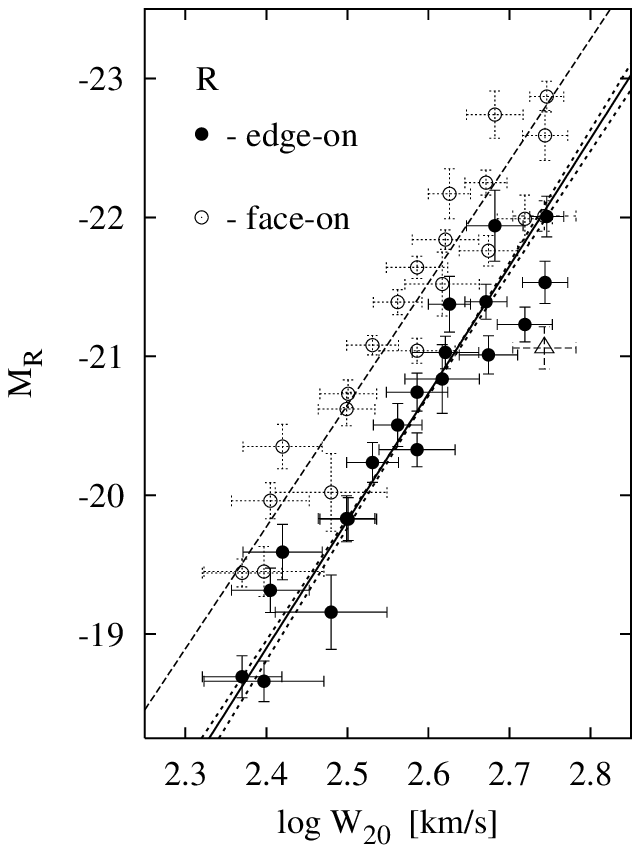}~
\hspace{-6cm}
\includegraphics[width=12cm,height=7cm]{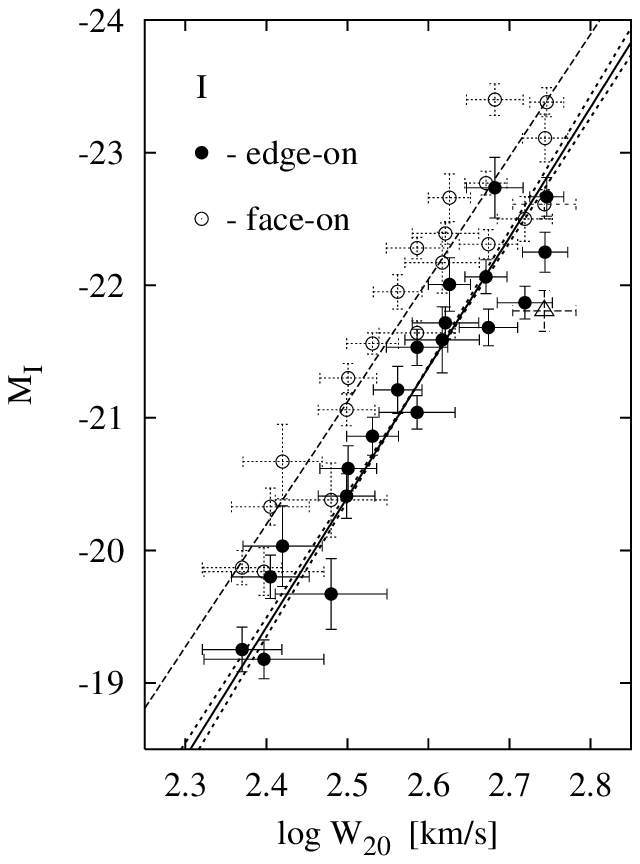}
\centering
\caption{R and I-band Tully-Fisher relations. The denotations are the same as in Fig. 1.}
\label{Fig. 2.}
\end{figure}
Both figures reveal that the application of the internal extinction model, which depend on the inclination correction, is of importance for the zero point values of the TF calibration relations. It is apparent that the two data sets on the diagrams are shifted  with approximately 1 mag.
	
	One strongly deviating point (the galaxy NGC 4414) was not taken into account when determining the fit parameters in all bands for the edge-on relations. In the upper fits, plotted from the calibration equations in \cite{SMH00}, that galaxy is not excluded.
	
	The errors of the edge-on magnitudes are calculated from the equation:

\begin{eqnarray}
\sigma^{2}(M_{e})&=&\sigma^{2}(m)+2.5c_{L}[\log(a/b)\delta(c_{L})]^{2}+[\delta(a/b)/\ln(10)]^{2}\nonumber\\
&+&[\log(q_{0})\delta(c_{L})]^{2}+\sigma^{2}(\mu)+\sigma^{2}(A)\nonumber
\end{eqnarray}
where $\delta(c_{L})=\sigma(c_{L})/c_{L}$ and $\mu=(m-M)_{0}$. The axis ratio errors were derived from the equation:
\[
\sigma(a/b)=\frac{\cos{i}\,\sin{i}(1-q_{0}^{2})}{[\cos^{2}{i}(1-q_{0}^{2})+q_{0}^{2}]\,^{3/2}}\sigma(i)\
\]
	This are computed without knowing the error of the parameter $q_{0}$. The inclinations errors are taken from \cite{MH00}. The $\log W_{20}$ errors were taken from \cite{SMH00}.

	The scatters of the TFRs in Fig.1 and 2 are due to the observational errors and physical differences of the galaxies. We obtained the following regressions:
	\begin{eqnarray}
	M_{B}&=&0.92(\pm 2.29)-7.87(\pm 0.89)\log W_{20},\;\;\sigma = 0.40 \mbox{ 			mag}\nonumber\\
	M_{V}&=&2.75(\pm2.15)-8.83(\pm0.83)\log W_{20},\;\;\sigma = 0.36 \mbox{ mag}.\nonumber\\
	M_{R}&=&3.11(\pm1.91)-9.17(\pm0.75)\log W_{20},\;\;\sigma = 0.31 \mbox{ mag}.\nonumber\\
	M_{I}&=&4.10(\pm2.08)-9.80(\pm0.82)\log W_{20},\;\;\sigma = 0.34 \mbox{ mag}\nonumber
	\end{eqnarray}
	Here the standard errors s correspond to relative distance errors of 18\%, 17\%, 14\% and 16\%, respectively. The error values in all bands are of one and the same order, but these in R and I band are smaller, probably due to a lower influence of the internal extinction at the larger wavelengths \cite{TP00}. For comparison, the values of s of the TFRs in the case of face-on view of the same galaxies are 0.43, 0.37, 0.34 and 0.36 mag. \cite{SMH00}.
	
	\section{Conclusions}
	In the present paper we calibrated the TFRs for the case of edge-on spiral galaxies in four optical bands. In our case, like in other studies \cite{KMK02,SMH00,SGG02}, the slope of the relations grows up with the increment of the wavelength and in R and I band it is close to the theoretical value - about  -10 \cite{AHM79}. In the same time the SE decreases, possibly due to the smaller internal extinction.

	Generally, the scatter and slope of the TFRs depends also on the completeness of the sample. This complicate problem is discussed in \cite{SMH00} and references therein. We believe that the considered sample is statistically representative and that it allows an accurate derivation of the TFR's parameters. Nevertheless, the SEs of the TFRs for the edge-on galaxies appears to be small enough and their usefulness as good distance indicators becomes obvious. It seems that the other version of the TFR - ''absolute diameter - HI line width'' may be confidently calibrated as well as \cite{N98}.

	According to the results of this work a special observing program should be carried on in order to derive more accurate distances for a larger sample of nearby edge-on galaxies that will lead to a better calibration of the TFR.\\

		\textbf{Acknowledgments}. This work is partially supported by the National Science Fund of the Bulgarian Ministry of Science and Education under grant No.1302/2003.

\end{document}